\documentclass[pre,twocolumn]{revtex4}
\usepackage{epsfig}
\usepackage{bm}
\usepackage{amsmath}
\usepackage{amssymb}

\begin{document}

\title{Spontaneous breaking of local mirror symmetry in superfluid and in BEC turbulence}

\author{A. Bershadskii}

\affiliation{
ICAR, P.O. Box 31155, Jerusalem 91000, Israel
}

\begin{abstract}

  There exist inherent mechanisms of spontaneous breaking of local mirror symmetry (parity) in fluid turbulence. A good example is Kerr's mechanism based on the inviscid vortex interactions. Such interactions can result in the appearance of the local adjacent regions with strong oppositely signed helicity, while the global (net) helicity still remains zero due to the global mirror symmetry. It is shown, using results of direct numerical simulations and laboratory measurements, that the spontaneous breaking of local mirror symmetry can dominate turbulent dynamics in the superfluid helium II and in the Bose-Einstein condensate turbulence. The notion of helical distributed chaos has been used for this purpose.  

\end{abstract}

\maketitle

\section{Introduction} 

   Spontaneous breaking of local mirror (reflectional) symmetry in the fluid dynamics is a long-known phenomenon as well as its relation to the helicity properties (see, for instance, Ref. \cite{bkt} and references therein). A good example of inherent mechanisms of spontaneous breaking of local mirror symmetry in inviscid fluid dynamics (related to turbulence) is Kerr's mechanism based on the inviscid vortex interactions \cite{ker} (see also Ref. \cite{hk}). Those inviscid interactions can result in the emergence of a pair of adjacent regions with strong helicity of the opposite sign. At these events, the global helicity still remains equal to zero due to the global mirror symmetry. One can expect that this and similar mechanisms can play a crucial role just in the superfluid turbulence.\\ 
   
    There are three main models of superfluid dynamics. The two-fluid HVBK model,  the Vortex Filament Model, and the Gross–Pitaevskii model. The classic-like HVBK model is a combination of the Navier-Stokes (for the normal component) and Euler (for the superfluid component) equations with a mutual coupling term in both equations. There is no problem with introducing classic helicity in this model (see for an excellent review Ref. \cite{holm}). The Vortex Filament Model is used for the mesoscale description of the superfluid dynamics of the helium II at the non-zero temperatures (these scales are typical for the present experiments with the helium II). It is also important that in this model a natural extension of the classical definition of helicity to the mesoscale superfluid dynamics is possible \cite{gal} (see also the Ref. \cite{hhs}). The Euler Vortex Filament Model is a reduced form of the Gross-Pitaevskii quantum model (with neglected variations of the wave function over the superfluid coherence length). The Gross-Pitaevskii quantum model is also useful for the description of the turbulence in the Bose-Einstein condensate (BEC). In the full Gross-Pitaevskii model there is a problem with introducing the helicity because both velocity and vorticity are singular on the quantized vortices centerlines. Therefore one needs in a smoothing/regularization for this purpose. Though, it was shown that sufficiently populated bundles of quantized vortices behave like the classical vortex tubes in a {\it viscous} fluid \cite{ked}. We will consider all these models in the present paper. \\
    
    We will use the notion of distributed chaos as the main theoretical tool. The distributed chaos is a parametric ensemble randomization of the deterministic chaos preserving the smoothness of the trajectories, that results in the stretched exponential spectra (violation of the smoothness would result in the scaling - power-law spectra). It will be shown that the adiabatic quasi-invariants related to the second and third moments of helicity distribution determine the distributed chaos stretched exponential energy spectra both for the superfluid and the Bose-Einstein condensate turbulence.

\section{Deterministic chaos in superfluid flows and in the BEC}

  In the recent paper Ref. \cite{gal} results of the numerical simulations with the Vortex Filament Model model have been reported. This model is used for the mesoscale description of the superfluid dynamics of helium II at non-zero temperatures.  \\
  
    Figure 1 shows the energy spectrum obtained for the thermally-driven counter flow of the superfluid He-II at the temperature T = 1.9 K (let us recall that the superfluid phenomenon appears below the temperature $T= 2.17$K). The spectral data were taken from Fig. 6 of the Ref. \cite{gal}. The dashed curve indicates the exponential spectral law
$$
 E(k) \propto \exp-(k/k_c)   \eqno{(1)}   
 $$ 
 where $k$ is the wavenumber. The position of the parameter $k_c$ is indicated in Fig. 1 by the dotted arrow. \\
 
   The exponential spectra (both temporal and spatial) are typical for the deterministic chaos with smooth trajectories (see, for instance, Refs. \cite{oh}-\cite{kds} and references therein). \\
   
   An Euler (incompressible) Vortex Filament Model of the superfluid dynamics was also studied numerically in the paper Ref. \cite{kiv} with initial conditions consisting of four superfluid vortex rings located on opposite sides of a cube and moving (due to their mutual orientation) toward the cube's center.  
   
   Figure 2 shows the power spectrum of the superfluid velocity obtained in this numerical simulation (the spectral data were taken from Fig. 2 of the Ref. \cite{kiv}), and again the dashed curve indicates the exponential spectral law Eq. (1). \\

%%%%%%%%%%%%%%% 1 %%%%%%%%%%%%%%%%%%
\begin{figure} \vspace{-1.3cm}\centering
\epsfig{width=.47\textwidth,file=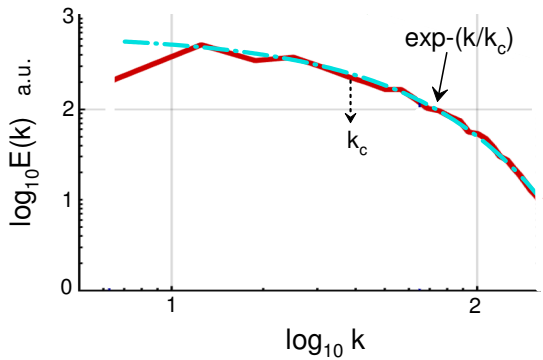} \vspace{-4.5cm}
\caption{Energy spectrum for the thermally-driven counter flow of the superfluid He-II at the temperature T = 1.9 K obtained with the Vortex Filament Model \cite{gal}.} 
\end{figure}
%%%%%%%%%%%%%%%%%%%%%%%%%%%%%%%%%%% 
 %%%%%%%%%%%%%%% 2 %%%%%%%%%%%%%%%%%%
\begin{figure} \vspace{-0.5cm}\centering
\epsfig{width=.47\textwidth,file=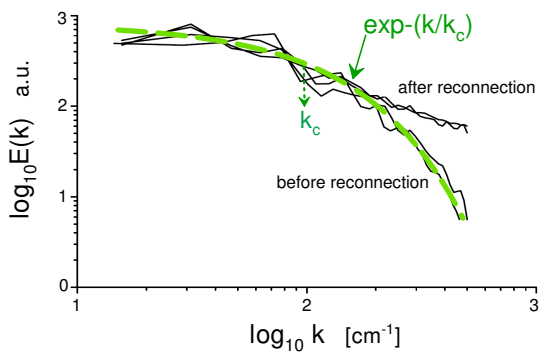} \vspace{-5cm}
\caption{Power spectrum of the superfluid velocity before and after reconnection for an Euler Vortex Filament Model \cite{kiv}.} 
\end{figure}
%%%%%%%%%%%%%%%%%%%%%%%%%%%%%%%%%%% 
%%%%%%%%%%%%%%% 3 %%%%%%%%%%%%%%%%%%
\begin{figure} \vspace{-1.5cm}\centering
\epsfig{width=.45\textwidth,file=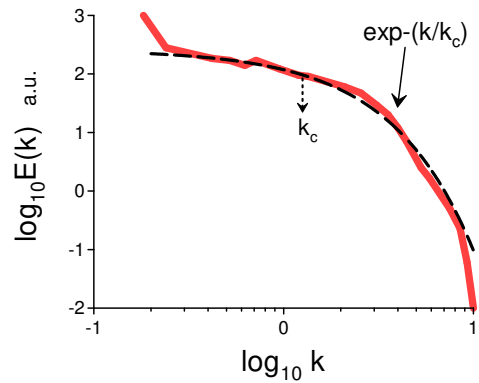} \vspace{-3.83cm}
\caption{Incompressible kinetic energy spectrum for the Gross-Pitaevskii quantum equation with a large-scale random potential \cite{kad}.} 
\end{figure}
%%%%%%%%%%%%%%%%%%%%%%%%%%%%%%%%%%% 
  The Euler Vortex Filament Model is a reduced form of the Gross-Pitaevskii quantum model (with neglected variations of the wave function over the superfluid healing length).

   In a recent paper \cite{kad} the full Gross-Pitaevskii equation 
$$
(i-\gamma) \hbar \frac{\partial \Psi}{\partial t} = - \frac{\hbar^2}{2m} {\nabla}^2 \Psi
		+ U({\bm r}, t)\Psi + \mathrm{g} | \Psi |^2 \Psi,    \eqno{(2)}
$$
was numerically solved with a random, homogeneous, and isotropic external potential $U({\bm r}, t)$ and the initial condition was taken as a homogeneous state. The characteristic spatial scale of the $U({\bm r}, t)$ was taken of the order of the spatial system size. In the Eq. (1) $\Psi ({\bf r}, t)$ is the macroscopic wave function, and $\gamma$ is a phenomenological dissipation constant. \\

   The superfluid velocity can be obtained by the Madelung's transformation $\psi({\bf x},t)=\sqrt{\frac{\rho({\bf x},t)}{m}}\exp{[i \frac{m}{\hbar}\phi({\bf x},t)]}$, where $\rho({\bf x},t)$ is the density of the superfluid and ${\bf v} = \nabla \phi$ is the velocity field (see below for more details). \\

  Figure 3 shows the incompressible kinetic energy spectrum obtained in this direct numerical simulation (the spectral data were taken from Fig. 4$c^{\prime\prime}$ of the Ref. \cite{kad}). The dashed curve indicates the exponential spectral law Eq. (1) typical for deterministic chaos, despite the random nature of the external potential  $U({\bm r}, t)$. In relation to this, it should be noted that there were strong correlations in the observed vorticity distributions.
   
 \section{Helicity moments} 
 
    The dynamics of He II in a first approximation can be described by the classic-like HVBK model: a combination of the Navier-Stokes (for the normal component) and Euler (for the superfluid component) equations
$$
\frac{D {\bf v}_n}{D t} = -\frac{1}{\rho_n} \nabla p_n  + \frac{\rho_s}{\rho} {\bf F}_{ns} 
                    + \nu \nabla^2 {\bf v}_n + {\bf f}_n^{ext}, \eqno{(3)}
 $$
 $$                   
\frac{D {\bf v}_s}{D t} = -\frac{1}{\rho_s} \nabla p_s   - \frac{\rho_n}{\rho}{\bf F}_{ns}
                    + {\bf f}_s^{ext},  \eqno{(4)}  
$$                     
and the incompressibility equations
$$
\nabla \cdot {\bf v}_n=0, ~~~~~\nabla \cdot {\bf v}_s=0     \eqno{(5-6)}
$$
 where the index $n$ refers to the normal fluid and the index $s$ refers to the superfluid, $\rho=\rho_n+\rho_s$ is the total density, the partial pressures are $p_n=(\rho_n/\rho)p+\rho_s ST$ and $p_s=(\rho_s/\rho) p -\rho_s ST$, with $p$, $T$, $S$ as pressure, temperature and specific entropy.

The mutual coupling term in the first order is \cite{srl}
$$
{\bf F}_{ns} = -\frac{B}{2} \vert  {\boldsymbol \omega}_s \vert {\bf v}_{ns}, \eqno{(7)}
$$
where ${\boldsymbol \omega}_s = [\nabla \times {\bf v}_s]$ is the superfluid vorticity, ${\bf v}_{ns} = {\bf v}_n - {\bf v}_s$, and $B$ is a constant. The external forcing terms are ${\bf f}_{n}^{ext}$ and ${\bf f}_{s}^{ext}$ .\\

  One can divide the spatial domain into a system of cells with spatial volumes $V_j$ (the index $j$ enumerates the cells) moving with the fluid, so that the superfluid vorticity is tangential at the boundary $S_j$ of each cell: ${\boldsymbol \omega}_s \cdot {\bf n}=0$. The moments of the superfluid helicity distribution $h_s({\bf r},t) = {\bf v}_s \cdot {\boldsymbol \omega}_s$ can be then defined on this system as \cite{lt},\cite{mt}
$$
I_n = \lim_{V \rightarrow  \infty} \frac{1}{V} \sum_j H_{j}^n  \eqno{(8)}
$$
where the superfluid helicity in the volume $V_j$ is
$$
H_j = \int_{V_j} h_s({\bf r},t) ~ d{\bf r}.  \eqno{(9)}
$$
and $V$ is the total spatial volume of the system. \\

  If the system has global mirror symmetry, then the global superfluid helicity $I_1$ and odd moments are identically equal to zero. There can be events of spontaneous breaking of the local mirror symmetry which result in the appearance of the cells with non-zero helicity, but the global superfluid helicity and the odd moments are still identically equal to zero. It is shown numerically in the paper Ref. \cite{ker}, for instance, that a spontaneous local breaking of the mirror symmetry can be due to the inviscid vortex interactions. Those interactions can result in the emergence of a pair of adjacent regions with strong helicity of the opposite sign. At these events, the global helicity (and the odd moments) still remains equal to zero due to the global mirror symmetry. It is also shown that this mechanism can be considered inherent to turbulent flows. \\
  
  The dynamic of the superfluid helicity of the helical cells in the HVBK model (with zero external forces) is described by equation \cite{holm}
 $$
 \frac{D H_j}{D t} =  \oint_{S_j} ({\bf n} \cdot {\bf v}_L) h_s({\bf r},t)  ~ dS \eqno{(10)}
$$
  Here ${\bf v}_L$ is the so-called velocity of the vortex line and the superfluid vorticity dynamics is described by the equation
 $$
 \frac{\partial   {\boldsymbol \omega}_s}{\partial t} = [\nabla \times ({\bf v}_L \times  {\boldsymbol \omega}_s)] \eqno{(11)}
 $$
  
  From the Eq. (10), one can see that the superfluid helicity in the cell $j$ will be conserved, provided ${\bf v}_L$ is tangential at the boundary of the cell or the helicity density $h_s({\bf r},t)$ is zero on the boundary of the cell.\\
   
 The second condition of the helicity $H_j$ conservation: $h_s({\bf r},t) =0$ on the cell's boundary, seems to be more appropriate for the spontaneous breaking of local mirror symmetry.\\ 
 
   If the external force $ {\bf f}_s^{ext}$ is concentrated in the large scales only, then despite the mean helicity cannot be generally considered as an invariant due to this force the higher moments of the superfluid helicity density $h_s({\bf r},t)$ are still quasi-invariants in this case. This is because the main contribution to the higher moments Eq. (8) comes from the cells with smaller characteristic scales \cite{bt}.
 
    It should be noted that the mutual coupling is most effective at the largest spatial scales, that also favors the conservation of the higher helicity moments. \\

   For the even values of $n$ the moments $I_n$ are finite (non-zero) quasi-invariants in a certain range of scales both for globally helical and non-helical flows (with the spontaneous breaking of local mirror symmetry) \cite{mt}. \\
   
   For the odd moments the situation is more complex. Let us denote the helicity of the cells with negative helicity $H_j^{-}$ and with positive helicity $H_j^{+}$, and denote
$$
I_n^{\pm} = \lim_{V \rightarrow  \infty} \frac{1}{V} \sum_j [H_{j}^{\pm}]^n  \eqno{(12)}
$$ 
where the summation is made for the negative (or positive) helicity cells only.  

   Due to the global mirror symmetry $I_n = I_n^{+} + I_n^{-} =0$ for the odd values of $n$. Consequently $I_n^{+} = - I_n^{-}$ for the odd values of $n$, and $|I_n^{\pm}|$ can be a finite (non-zero) quasi-invariant at a certain range of scales for the flows with spontaneously broken local mirror symmetry.

\section{Distributed chaos }

\subsection{Third moment of the helicity distribution}

   Let us turn to the description of the chaotic/turbulent flows of the superfluid with the spontaneous breaking of the mirror symmetry. If the characteristic scale $k_c$ in the exponential spectrum Eq. (1) fluctuates to evaluate the kinetic energy spectrum one should use a kind of the ensemble averaging
$$
E(k) \propto \int_0^{\infty} P(k_c) \exp -(k/k_c)dk_c \eqno{(13)}
$$    
where the probability distribution $P(k_c)$ plays a key role. \\

   To find the probability distribution $P(k_c)$ for the superfluid chaotic/turbulent flow with spontaneously broken local mirror symmetry let us use the $|I_3^{\pm}|$ quasi-invariant, for simplicity. The characteristic velocity $v_c$ can be related to the characteristic scale $k_c$ using the dimensional considerations
 $$
 v_c \propto |I_3^{\pm}|^{1/6}~ k_c^{1/2}    \eqno{(14)}
$$       
 
   Assuming normal (Gaussian) distribution of the characteristic velocity $v_c$ (with zero mean) (see, for instance, Ref. \cite{my} and references therein), one can readily find from the Eq. (14) 
$$
P(k_c) \propto k_c^{-1/2} \exp-(k_c/4k_{\beta})  \eqno{(15)}
$$
here $k_{\beta}$ is a constant parameter.

%%%%%%%%%%%%%%% 4 %%%%%%%%%%%%%%%%%%
\begin{figure} \vspace{-1.5cm}\centering
\epsfig{width=.45\textwidth,file=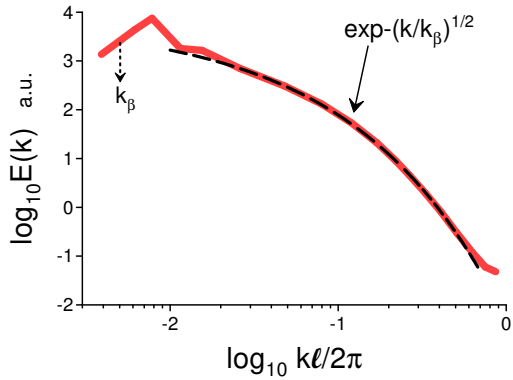} \vspace{-4cm}
\caption{Superfluid energy spectrum for the truncated HBVK simulation at $T \simeq 2.16$K. } 
\end{figure}
%%%%%%%%%%%%%%%%%%%%%%%%%%%%%%%%%%% 
%%%%%%%%%%%%%%% 5 %%%%%%%%%%%%%%%%%%
\begin{figure} \vspace{-0.37cm}\centering
\epsfig{width=.47\textwidth,file=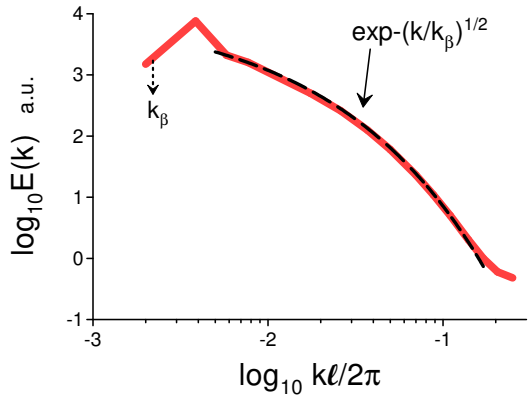} \vspace{-4.1cm}
\caption{  As in Fig. 4 but for $T = 1.15$K.} 
\end{figure}
%%%%%%%%%%%%%%%%%%%%%%%%%%%%%%%%%%%
%%%%%%%%%%%%%%% 6 %%%%%%%%%%%%%%%%%%
\begin{figure} \vspace{-1.63cm}\centering
\epsfig{width=.45\textwidth,file=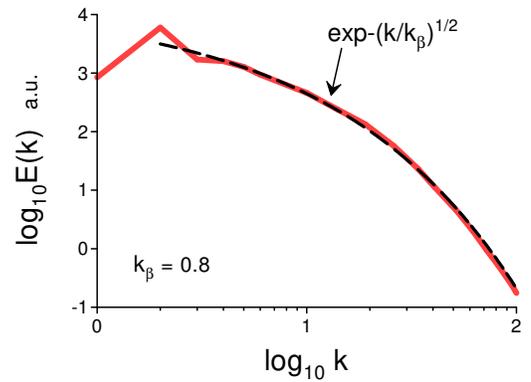} \vspace{-3.9cm}
\caption{Superfluid energy spectrum for the HBVK simulation with $\nu_n/\nu_s =4$ at $T \simeq 2.16$K. } 
\end{figure}
%%%%%%%%%%%%%%%%%%%%%%%%%%%%%%%%%%% 
%%%%%%%%%%%%%%% 7 %%%%%%%%%%%%%%%%%%
\begin{figure} \vspace{-0.3cm}\centering
\epsfig{width=.45\textwidth,file=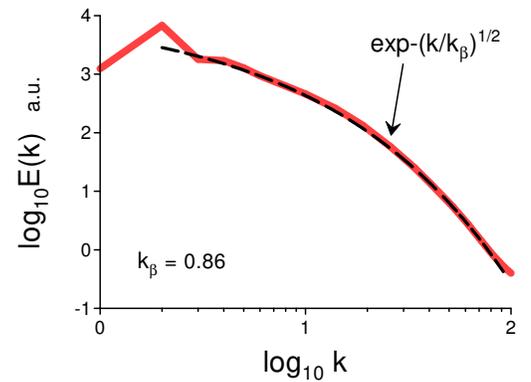} \vspace{-3.8cm}
\caption{  As in Fig. 6 but for $\nu_n/\nu_s =100$} 
\end{figure}
%%%%%%%%%%%%%%%%%%%%%%%%%%%%%%%%%%%

    Substituting the probability distribution Eq. (15) into Eq. (13) one obtains
$$
E(k) \propto \exp-(k/k_{\beta})^{1/2}  \eqno{(16)}
$$     
 
 \subsection{The HVBK models}
 
  In the paper Ref. \cite{srl} (see also Ref. \cite{sclr}) results of direct numerical simulations of the HVBK model Eqs. (3-7) were reported. The DNS were performed in a cubic domain with periodic boundary conditions. A large scale random forcing (acting for $1.5 < |{\bf k}| < 2.5$) was  imposed on the normal component at the temperature $T \simeq 2.16$K ($\rho_s/\rho_n = 0.1$ ) and on the superfluid component at the temperature $T =1.15$K ($\rho_s/\rho_n = 40$). The computations were truncated at the the intervortex scale ${\ell }$. \\
  
    Figure 4 and 5 show the energy spectra obtained at $T \simeq 2.16$K and $T =1.15$K correspondingly (the spectral data were taken from Fig. 3a of the Ref. \cite{blr}). The dashed curves in Figs. 4 and 5 are drawn to indicate the stretched exponential spectral decay Eq. (16). The dotted arrows indicate the position of the scale $k_{\beta}$.\\

    In another version of the HBVK model \cite{rbl} a small scale dissipation of the superfluid component is also taken into account by introducing an effective viscosity term
$$
\frac{D {\bf v}_n}{D t} = -\frac{1}{\rho_n} \nabla p_n  + \frac{\rho_s}{\rho} {\bf F}_{ns} 
                    + \nu_n \nabla^2 {\bf v}_n + {\bf f}_n^{ext}, \eqno{(17)}
 $$
 $$                   
\frac{D {\bf v}_s}{D t} = -\frac{1}{\rho_s} \nabla p_s   - \frac{\rho_n}{\rho}{\bf F}_{ns} + \nu_s \nabla^2 {\bf v}_s
                    + {\bf f}_s^{ext},  \eqno{(18)}  
$$              
The artificial viscosity term $\nu_s \nabla^2 {\bf v}_s$ was suggested in order to take into account the main small scale effects that the HVBK model Eqs. (3-4) cannot directly resolve (e.g. Kelvin waves and quantized vortex reconnections \cite{kl},\cite{bpsl}\cite{vpk}, see below for more details).\\

 In the Ref. \cite{rbl} results of direct numerical simulations, using the Eqs. (17-18) (with the Eqs. (5-6)) in a cubic box with periodic boundary conditions, were reported.  The mutual coupling term in the first order was taken in the same form as for the previous HBVK simulation. The large scale random forcing (acting for $1.5 < |{\bf k}| < 2.5$) was  imposed on the normal component at the temperature $T \simeq 2.16$K ($\rho_s/\rho_n = 0.1$ ) and $\nu_n/\nu_s = 4$ for one run and $\nu_n/\nu_s = 100$ for another run. \\
    
      Figures 6 and 7  show the energy spectra obtained at $\nu_n/\nu_s = 4$ and $\nu_n/\nu_s = 100$ correspondingly (the spectral data were taken from Fig. 2b of the Ref. \cite{rbl}). The dashed curves in Figs. 6 and 7 are drawn to indicate the stretched exponential spectral decay Eq. (16). One can also see that despite the large difference in the relation $\nu_n/\nu_s $ the energy spectra are slightly different in the value of $k_{\beta}$ only.\\

\subsection{Gross–Pitaevskii model}

%%%%%%%%%%%%%%%8 %%%%%%%%%%%%%%%%%%
\begin{figure} \vspace{-1.2cm}\centering \hspace{-2cm}
\epsfig{width=.57\textwidth,file=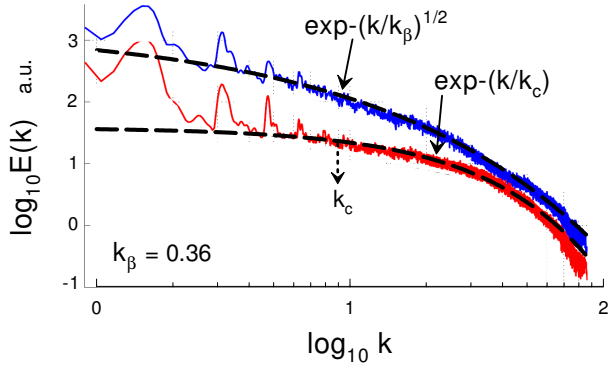} \vspace{-6.3cm} 
\caption{ Incompressible part of kinetic energy spectrum for different times of the evolution: $t =20$ (upper curve) and $t=50$ (lower curve).}
\end{figure}
%%%%%%%%%%%%%%%%%%%%%%%%%%%%%%%%%%%
    
 %%%%%%%%%%%%%%% 9 %%%%%%%%%%%%%%%%%%
\begin{figure} \vspace{-0.75cm}\centering \hspace{-1cm}
\epsfig{width=.5\textwidth,file=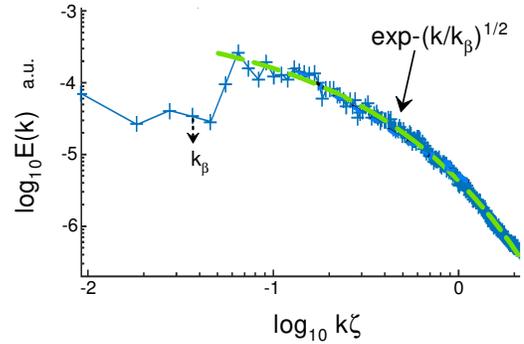} \vspace{-5.1cm}
\caption{Incompressible part of the kinetic energy spectrum computed at the distance $z \simeq 360\xi$ from the grid} 
\end{figure}
%%%%%%%%%%%%%%%%%%%%%%%%%%%%%%%%%%%
 %%%%%%%%%%%%%%% 10 %%%%%%%%%%%%%%%%%%
\begin{figure} \vspace{-0.45cm}\centering \hspace{-1cm}
\epsfig{width=.5\textwidth,file=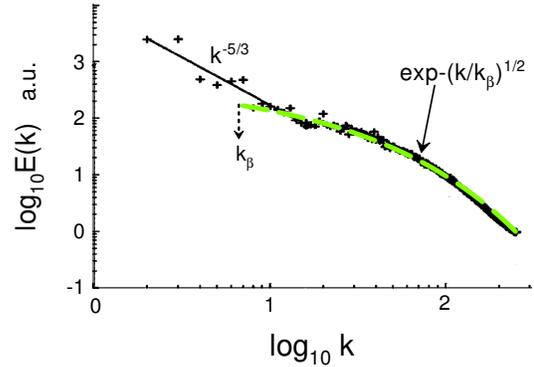} \vspace{-5.55cm}
\caption{Incompressible part of the kinetic energy spectrum of decaying Gross–Pitaevskii turbulence in the geometry of the Taylor-Green (vortex) flow.} 
\end{figure}
%%%%%%%%%%%%%%%%%%%%%%%%%%%%%%%%%%%

The Gross–Pitaevskii model of the superfluids turned out to be a good first-order approximation for the superfluid $^4He$ (when the normal fluid component is negligible) and for the Bose-Einstein condensate. \\

    Let us consider a simple dimensionless form of the Gross–Pitaevskii equation 
$$
2i \frac{\partial \psi}{\partial t}=-\nabla^2 \psi +\left ( |\psi|^2-1\right )\psi, \eqno{(19)}
$$
where the coherence length $\xi$ was taken as a unit of the spatial scale ($\xi$ is of the order of the quantized vortex core size). \\

    Using the Madelung's transformation $\psi = \sqrt{\rho} e^{i \theta}$ \cite{mad} one can map Eq. (19) into a classic like equation for a compressible fluid \cite{bag},\cite{ya},\cite{sal1}
$$
\rho \left(\frac{\partial v_i}{\partial t}+v_j\frac{\partial v_i}{x_j} \right) =
-\frac{\partial p}{\partial x_i} + \frac{\partial \tau_{ij}}{\partial x_j}  \eqno{(20)}
$$
Here $\rho=|\psi|^2$, ${\bf v}=\bm\nabla{\theta}$, $p=\rho^2/4$ and 
$$
\tau_{ij}=\frac{\rho}{4} \frac{\partial^2 \ln\rho}{\partial x_i \partial x_j} \eqno{(21)})
$$

    The $\tau_{ij}$ can be considered as a quantum stress and it is responsible for the reconnections of the quantized vortices. At length scales larger than the coherence length $\xi$ the quantum stress term is negligible in comparison to the pressure term and we obtain the Euler equation so that the helicity is approximately conserved for these scales. \\

    It should be noted that there is uncertainty with the definition of helicity in the full Gross–Pitaevskii model: both vorticity and velocity fields are singular on the quantized vortices centerlines. Though, it was shown that sufficiently populated bundles of quantized vortices behave like the classical vortex tubes in a viscous fluid \cite{ked}. Therefore, similarly to the classic viscous fluids, the helicity (and its moments) can be still considered as an adiabatic invariant in an inertial range of scales. This also justifies the appearance of the viscous term $\nu_s \nabla^2 {\bf v}_s$ in the HVBK model Eq. (18). \\
    
    In the paper Ref. \cite{hhg} results of a direct numerical simulation of chaotic/turbulent states in the Bose-Einstein condensate (BEC) with the Gross–Pitaevskii model were reported. In this simulation, the chaotic/turbulent states were generated by the bending-wave instabilities of a single-vortex ring placed inside the BEC.  At a certain stage of the instability development the vortex ring is breaking into numerous vortex filaments which are tangled, reconnected, and continuously deformed. The interaction between the vortex filaments results in their tendency for localization in the outer region of the condensate. At a later stage the vortex filaments are gradually disappeared while the density isosurface is progressively wrinkled. \\
    
    The superfluid kinetic energy can be considered as a sum of the incompressible (vortices) and compressible (sound waves) components. Figure 8 shows the incompressible part of kinetic energy spectrum for different times of the evolution: $t =20$ (upper curve) and $t=50$ (lower curve). The spectral data were taken from Fig. 3 of the Ref. \cite{hhg}. The dashed curves are drawn to indicate the spectral law Eq. (16) for $t=20$ and the spectral law Eq. (1) for $t = 50$ (the spectrum for $t=20$ was multiplied by factor 10 in comparison to that for $t=50$, for more clear visual separation). One can see that for $t =20$ the spectrum indicates the distributed chaos whereas for $t=50$ the spectrum indicates the deterministic chaos (cf Fig. 3).  \\

    In the paper Ref. \cite{kru} a superfluid flow behind a grid was numerically simulated using the modified Gross-Pitaevskii equation in a box with periodic boundary conditions
$$
i\hbar\left(\frac{\partial \psi}{\partial t}+\bf{v_0}\cdot\nabla\psi\right)=- \frac{\hbar^2}{2m}\nabla^2 \psi + g\,|\psi|^2\psi -\tilde{\mu}\, \psi+V_{\rm grid}({\bf x})\,\psi  \eqno{(22)}
$$
where $\bf{v_0}\cdot\nabla\psi$ is the advection term with the mean velocity $\bf{v_0}$, and the grid is simulated by a repulsive potential $V_{\rm grid}({\bf x})$. The $\tilde{\mu}$ is the chemical potential.\\

   Figure 9 shows the incompressible part of the kinetic energy spectrum computed at the distance $z \simeq 360\xi$ from the grid.  The spectral data were taken from Fig. 2c of the Ref. \cite{kru}. The dashed curve is drawn to indicate the spectral law Eq. (16). \\

   In the paper Ref. \cite{nab} decaying superfluid turbulence in the geometry of the Taylor-Green (vortex) flow \cite{bra} 
$$ 
{\bf v}(x, y, z) = v_0 \left[ (\sin x \cos y \cos z) {\bf\hat{e}_x} - (\cos x \sin y \cos z)\bf{\hat{e}_y} \right] \eqno{(23)}
$$   
   was numerically simulated using the Gross-Pitaevskii equation in a box with periodic free-slip boundary conditions. It should be noted that the global (net) helicity of the Taylor-Green flow is zero. \\
   
   Figure 10 shows the incompressible part of the kinetic energy spectrum at $t = 5.5$ in the terms of the Ref. \cite{bra}. The spectral data were taken from Fig. 2b of the Ref. \cite{nab}. The dashed curve is drawn to indicate the spectral law Eq. (16). \\

\subsection{A laboratory experiment}

   In papers Refs. \cite{roch},\cite{sal2} results of the experiments in the N\'{E}EL wind tunnel with the He-II were reported. In this experiment a He-II loop was confined in a cylindrical cryostat and a steady flow of He-II was powered by a centrifugal pump (see Fig. 11). A honeycomb breaks the spin motion generated by the centrifugal pump. The temperature of the He-II was regulated near 1.55-1.60K (that corresponds to $\rho_s/\rho = 86-84\%$ of the two-fluid model). The mean velocity at the probe's location was varied from 0.6m/c to 1.3 m/s.\\
   
   There is a problem to interpret measurements of the velocity fluctuations in He-II. The most direct and simple interpretation can be made for the measurements by probes similar to the Pitot tube. However, this probe has low resolution. Therefore only the results made for the low-frequency range of the scales are reliable for these measurements. Another problem with these measurements is that they can be directly (linearly) interpreted for small mean velocities only (for large mean velocities the second-order corrections from quadratic velocity fluctuations and static pressure fluctuations result in significant bias).  \\

%%%%%%%%%%%%%%% 11 %%%%%%%%%%%%%%%%%%
\begin{figure} \vspace{-0.45cm}\centering  \hspace{-1.5cm}
\epsfig{width=.45\textwidth,file=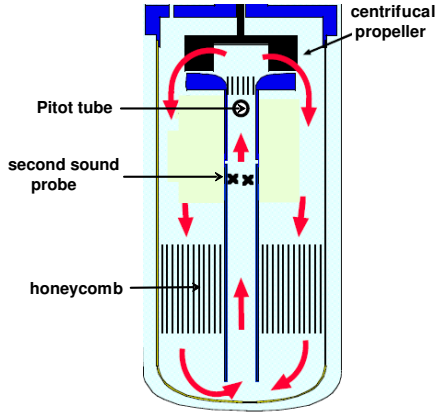} \vspace{-3.8cm}
\caption{Scheme of the He-II loop confined in a cylindrical cryostat and a flow of He-II powered by a centrifugal pump.} 
\end{figure}
%%%%%%%%%%%%%%%%%%%%%%%%%%%%%%%%%%%      
 %%%%%%%%%%%%%%% 12 %%%%%%%%%%%%%%%%%%
\begin{figure} \vspace{-0.6cm}\centering \hspace{-1cm}
\epsfig{width=.48\textwidth,file=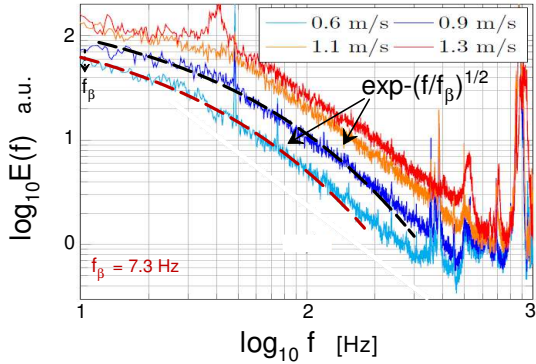} \vspace{-5cm}
\caption{Power spectra of velocity fluctuations measured by the stagnation pressure probe (the Pitot tube, Fig. 11) at different mean velocities and the temperature $T= 1.55$K.} 
\end{figure}
%%%%%%%%%%%%%%%%%%%%%%%%%%%%%%%%%%%
 %%%%%%%%%%%%%%% 13 %%%%%%%%%%%%%%%%%%
\begin{figure} \vspace{-0.5cm}\centering
\epsfig{width=.48\textwidth,file=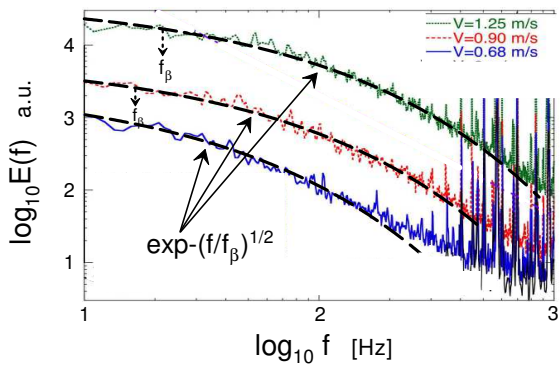} \vspace{-5.55cm}
\caption{ Power spectra of the vortex line density measured by the second-sound probe at the temperature $T = 1.6$K in the same laboratory installation (Fig. 11).} 
\end{figure}
%%%%%%%%%%%%%%%%%%%%%%%%%%%%%%%%%%% 

   Figure 12 shows power spectra of velocity fluctuations measured by the stagnation pressure probe (the Pitot tube, Fig. 11) at different mean velocities and the temperature $T= 1.55$K. The spectral data were taken from Fig. 8 of the Ref. \cite{sal2}. The observed frequency power spectra can be related to the wavenumber power spectra using the Taylor's `frozen-in' hypothesis
 $E(f) = E(k)~2\pi/U$   (where $U$ is the mean velocity of the flow) and $f =Uk/2\pi$ (see, for instance, Ref. \cite{kv} and references therein). The dashed curves are drawn to indicate the spectral law Eq. (16). \\
   
  In a fully random superfluid vortex tangle the vortex line density is expected to be proportional to the corresponding kinetic energy (see, for instance, Ref. \cite{don}). On the other hand, the probability distribution of the characteristic scale $k_c$ for the vortex line density can behave similarly to the probability density of the characteristic scale for the velocity fluctuations. Therefore, one can expect that the power spectra of the velocity field and the vortex line density has similar form in the case of distributed chaos. \\
  
  Figure 13 shows the power spectra of the vortex line density measured by the second-sound probe at the temperature $T = 1.6$K in the same laboratory installation (Fig. 11). The spectral data were taken from Fig. 4 of the Ref. \cite{roch}. The dashed curves are drawn to indicate the spectral law Eq. (16) (using the Taylor's hypothesis).\\
  
  Finally, it should be noted that in this experiment the characteristic spatial scale of quantum effects is much lower than the resolution of the probes, i.e. the quasi-classic approximation (see the previous subsection and a recent review Ref. \cite{skr}) can be well applied in this case.

 \section{Second moment of helicity distribution}   

\subsection{Stretched exponential spectrum}
 
    Since for the flows with global mirror symmetry the odd moments of helicity distribution are equal to zero we have introduced the non-zero variable $|I_n^{\pm}|$ (see (Eq. (12)) to analyze the spontaneous breaking of the local mirror symmetry. Unlike the odd moments, the even moments of helicity distribution are non-zero in the case of spontaneous breaking of the local mirror symmetry even when the flow has global mirror symmetry \cite{mt}. Therefore, one can also use the $I_n$ with even values of $n$ for this analysis. \\
    
    The estimation Eq. (14) can be replaced by the estimation
$$
 v_c \propto |I_n|^{1/2n}~ k_c^{\alpha_n},    \eqno{(24)}
 $$    
here
$$
\alpha_n = 1-\frac{3}{2n}  \eqno{(25)},
$$  
and the stretched exponential spectrum Eq. (16) can be generalized as
$$
E(k) \propto \int_0^{\infty} P(k_c) \exp -(k/k_c)dk_c \propto \exp-(k/k_{\beta})^{\beta} \eqno{(26)}
$$  
 This stretched exponential generalization corresponds to the smoothness of the dynamics at the distributed chaos. Then $P(k_c)$ can be estimated from Eq. (26) for large $k_c$  \cite{jon}
$$
P(k_c) \propto k_c^{-1 + \beta/[2(1-\beta)]}~\exp(-\gamma k_c^{\beta/(1-\beta)}) \eqno{(27)}
$$     
(where the $\gamma$ is a constant).\\

    If $v_c$ has Gaussian distribution \cite{my} a relationship between the exponents $\beta_n$ and $\alpha_n$ can be readily obtained from the Eqs. (24) and (27)
$$
\beta_n = \frac{2\alpha_n}{1+2\alpha_n}  \eqno{(28)}
$$
 
  Substituting $\alpha_n $  from the Eq. (25) into the Eq. (28) one obtains
 $$
 \beta_n = \frac{2n-3}{3n-3}   \eqno{(29)}  
 $$
and for the value $n=2$ ($I_2$ is the Levich-Tsinober invariant \cite{lt})
$$
E(k) \propto \exp-(k/k_{\beta})^{1/3}  \eqno{(30)}
$$

\subsection{The HBVK model }
    
   For the HBVK model with an artificial viscosity term $\nu_s \nabla^2 {\bf v}_s$ Eqs. (17-18) \cite{rbl} (suggested in order to take into account the main quantum small scale effects)  the direct numerical simulation at the high superfluid temperature  $T \simeq 2.16$K and $\nu_n/\nu_s = 4 $ provides the superfluid energy spectrum which is well approximated by the stretched exponential spectrum Eq. (16) (see Fig. 6). However, for a low superfluid temperature $T \simeq 1.44$K ($\nu_n/\nu_s = 4 $)  the superfluid energy spectrum is well approximated by the stretched exponential spectrum Eq. (30) - Fig. 14. The spectral data were taken from Fig. 2b of the Ref. \cite{rbl}. The dashed curve in the Fig. 14 is drawn to indicate the spectral law Eq. (30). \\

 %%%%%%%%%%%%%%% 14 %%%%%%%%%%%%%%%%%%
\begin{figure} \vspace{-1.4cm}\centering
\epsfig{width=.45\textwidth,file=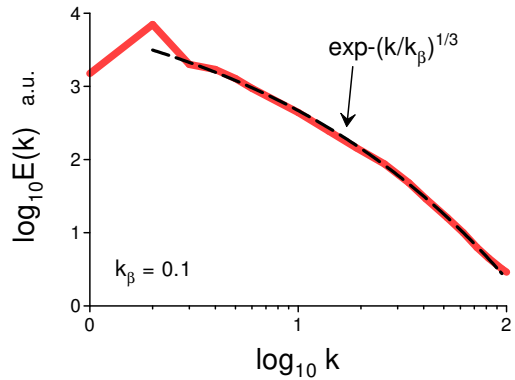} \vspace{-3.9cm}
\caption{The same as in Fig. 6 but for $T \simeq 1.44$K. } 
\end{figure}
%%%%%%%%%%%%%%%%%%%%%%%%%%%%%%%%%%%  
 %%%%%%%%%%%%%%% 15 %%%%%%%%%%%%%%%%%%
\begin{figure} \vspace{-0.5cm}\centering
\epsfig{width=.45\textwidth,file=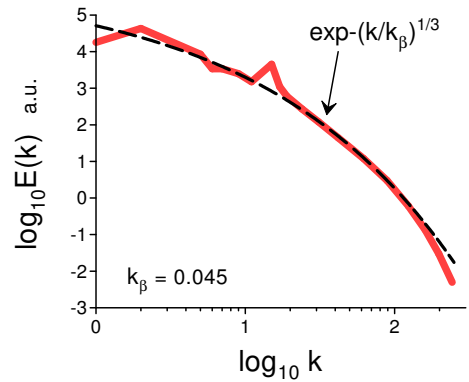} \vspace{-3.55cm}
\caption{Total energy spectrum computed at $T = 1.9$K, under strong counterflow, at the final time of the DNS.} 
\end{figure}
%%%%%%%%%%%%%%%%%%%%%%%%%%%%%%%%%%%
 %%%%%%%%%%%%%%% 16 %%%%%%%%%%%%%%%%%%
\begin{figure} \vspace{-0.5cm}\centering
\epsfig{width=.45\textwidth,file=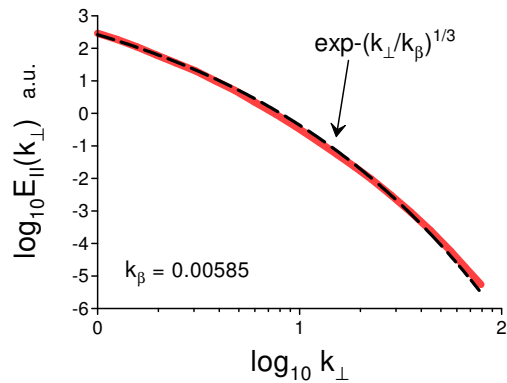} \vspace{-3.55cm}
\caption{The energy spectrum for the superfluid velocity fluctuations component parallel to the counterflow at $T = 1.65$K.} 
\end{figure}
%%%%%%%%%%%%%%%%%%%%%%%%%%%%%%%%%%%  
     In a recent paper Ref. \cite{pk} results of a direct numerical simulation using the HBVK model under strong counterflow were reported. In this case, the mean relative velocity between the normal and superfluid components ${\bf V}_{ns} ={\bf V}_n -  {\bf V}_s$ is non-zero, that can be readily taken into account in the Eqs. (17-18). The random external forces were taken Gaussian and zero-mean. The external forces were localized at the wavenumber $k_f =15$. The mutual friction forces were taken as ${\bf f}_s^{ext} = -(\rho_n/\rho_s){\bf f}_n^{ext} =\alpha\Omega_0 ({\bf v}_n - {\bf v}_s)$. At $t=0$ the velocity fluctuations ${\bf v}_n = {\bf v}_s = 0$. The DNS was performed in a 3D cubic and periodic box. \\
     
     Figure 15 shows the total energy spectrum computed in the Ref. \cite{pk} at $T = 1.9$K ($\rho_s/\rho_n =1.35$), under strong counterflow, at the final time of the DNS. The spectral data were taken from Fig. 1 of the Ref. \cite{pk}. The dashed curve in the Fig. 15 is drawn to indicate the spectral law Eq. (30). \\
     
 The energy spectra in the superfluid counterflows are strongly anisotropic. The turbulent velocity fluctuations are dominated by the component, parallel to the counterflow velocity
${\bf V}_{ns}$. Figure 16 shows the energy spectrum for the superfluid velocity fluctuations component parallel to the counterflow. The spectral data were taken from Fig. 6a of the Ref. \cite{bif}. The corresponding numerical simulations were performed at $T =1.65$K and the random external forces were localized at the wavenumber range $0.5 < k_f < 1.5$. The dashed curve is drawn to indicate the spectral law Eq. (30). \\
 
 \subsection{Bose-Einstein condensate turbulence with external current}    
 
 %%%%%%%%%%%%%%% 17%%%%%%%%%%%%%%%%%%
\begin{figure} \vspace{-1.6cm}\centering \hspace{-1.3cm}
\epsfig{width=.54\textwidth,file=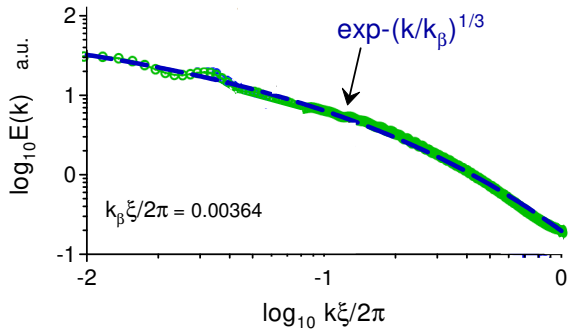} \vspace{-5.6cm}
\caption{Power spectrum of the velocity fluctuations obtained in the BEC DNS for the small-scale external currents.} 
\end{figure}
%%%%%%%%%%%%%%%%%%%%%%%%%%%%%%%%%%%     
%%%%%%%%%%%%%%% 18 %%%%%%%%%%%%%%%%%%
\begin{figure} \vspace{-1.3cm}\centering \hspace{-1cm}
\epsfig{width=.49\textwidth,file=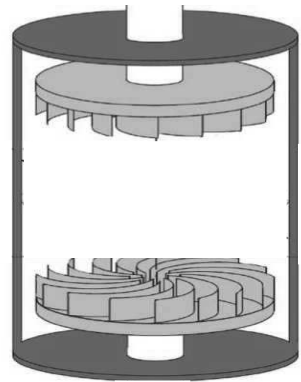} \vspace{-4.6cm}
\caption{A simplified sketch of the experimental setup.} 
\end{figure}
%%%%%%%%%%%%%%%%%%%%%%%%%%%%%%%%%%%
%%%%%%%%%%%%%%% 19%%%%%%%%%%%%%%%%%%
\begin{figure} \vspace{0.5cm}\centering
\epsfig{width=.45\textwidth,file=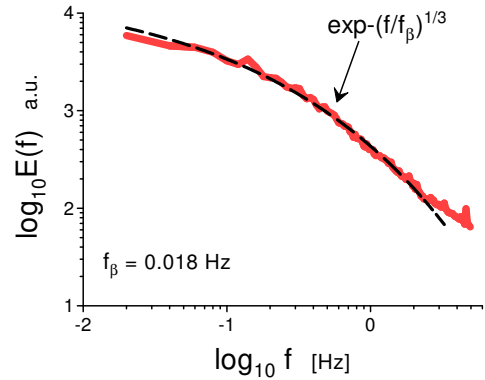} \vspace{-4cm}
\caption{The velocity power spectrum obtained by the miniature Pitot probe at $T= 2.0$K.} 
\end{figure}
%%%%%%%%%%%%%%%%%%%%%%%%%%%%%%%%%%%

  In paper Ref. \cite{ku} results of a direct numerical simulation of Bose-Einstein condensate turbulence with an external current were reported. The DNS was performed in the periodic box with the size $L =512 \xi$. 
  
 The injection of energy was made using the non-uniform gauge 
$$
\nabla \to \nabla - i M / (2 \hbar) {\bf v}_{\rm ext} \eqno{(31)}
$$
where  $\bf{v}_{\rm ext}$ is an external current and $M$ is the atomic mass. The external current was taken as the superposition of waves
$$
{\bf v}_{\rm ext}({\bf x}) = {\bf v}_0 \sum_{0 \leq |{\bf n}| \leq n_{\rm {cut}}} \cos( 2 \pi {\bf n} \cdot {\bf x} / L + \theta_{\bf{n}} ({\bf x} )) \eqno{(32)}
$$
 The random phase $\theta_{\bf{n}} ({\bf x} )$ is uniformly distributed for each ${\bf x}$ and ${\bf n}$ over $[0, 2 \pi)$. The external current with $ n_{\rm {cut}} =32$ was considered as a small-scale one.

  Dissipation was simulated using the cutoff wavenumber $k_{\rm c} = 2 \pi / \xi$ for truncation of the wave function. \\

    Figure 17 shows the power spectrum of the velocity fluctuations obtained in the DNS for the small-scale external currents. The spectral data were taken from Fig. 3 of the Ref. \cite{ku} (the scalar case). The ensemble average over 200 steady states (with different external currents) was taken in order to compute the spectrum. The dashed curve in the Fig. 17 is drawn to indicate the spectral law Eq. (30). \\

\subsection{A laboratory experiment}

%%%%%%%%%%%%%%% 20 %%%%%%%%%%%%%%%%%%
\begin{figure} \vspace{-1.1cm}\centering
\epsfig{width=.45\textwidth,file=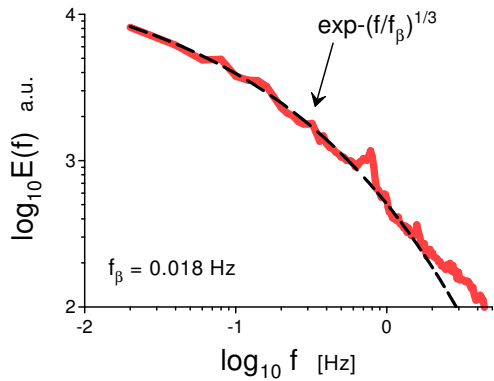} \vspace{-4cm}
\caption{The velocity power spectrum obtained by the miniature Pitot probe at $T= 1.58$K.} 
\end{figure}
%%%%%%%%%%%%%%%%%%%%%%%%%%%%%%%%%%%     

   In a recent paper \cite{salort} results of a laboratory experiment with the superfluid He-II in a large vessel were reported. The flow of the He-II was produced in a gap between two co-rotating at nearly the same angular speed disks with blades (see Fig. 18). The co-rotating disks produced a large vortex inside the spatial gap. At mid-height the azimuthal velocity of the fluid in the main body of the vortex was similar to that in solid-body rotation. However, close to the walls the azimuthal velocity was about independent of the $r$ (the viscous boundary layer near the wall was vanishingly small). A miniature Pitot probe was located at mid-height in the region with the constant (on $r$) azimuthal velocity. \\
   
    Figures 19 and 20 show the velocity power spectra obtained by the miniature Pitot probe in a `fully turbulent regime' at $T= 2.0$K and $T =1.58$K respectively. The spectral data were taken from the database https://$turbase.cineca.it/init/routes/\#/logging/\\
view\_dataset/34/tabfile$. The dashed curves in the Figs. 19 and 20 are drawn to indicate the spectral law Eq. (30) (with the application of the Taylor's frozen-in hypothesis). \\

 \section{Acknowledgement}

I thank  E. Levich for numerous discussions and the authors of the Ref. \cite{salort} for sharing their data.

\end{document}